\documentclass[]{article}
\voffset= - 1.5in
\hoffset= - 1.0in         

\def\d{\partial}

\def\bea{\begin{eqnarray}}
\def\eea{\end{eqnarray}}

\def\beq{\begin{equation}}
\def\eeq{\end{equation}}
\def\ba{\beq\new\begin{array}{c}}
\def\ea{\end{array}\eeq}
\def\be{\ba}
\def\ee{\ea}

\makeatletter
\newdimen\normalarrayskip              
\newdimen\minarrayskip                 
\normalarrayskip\baselineskip
\minarrayskip\jot
\newif\ifold             \oldtrue            \def\new{\oldfalse}
\def\arraymode{\ifold\relax\else\displaystyle\fi} 
\def\eqnumphantom{\phantom{(\theequation)}}     
\def\@arrayskip{\ifold\baselineskip\z@\lineskip\z@
     \else
     \baselineskip\minarrayskip\lineskip2\minarrayskip\fi}
\def\@arrayclassz{\ifcase \@lastchclass \@acolampacol \or
\@ampacol \or \or \or \@addamp \or
   \@acolampacol \or \@firstampfalse \@acol \fi
\edef\@preamble{\@preamble
  \ifcase \@chnum
     \hfil$\relax\arraymode\@sharp$\hfil
     \or $\relax\arraymode\@sharp$\hfil
     \or \hfil$\relax\arraymode\@sharp$\fi}}
\def\@array[#1]#2{\setbox\@arstrutbox=\hbox{\vrule
     height\arraystretch \ht\strutbox
     depth\arraystretch \dp\strutbox
     width\z@}\@mkpream{#2}\edef\@preamble{\halign
\noexpand\@halignto
\bgroup \tabskip\z@ \@arstrut \@preamble \tabskip\z@ \cr}%
\let\@startpbox\@@startpbox \let\@endpbox\@@endpbox
  \if #1t\vtop \else \if#1b\vbox \else \vcenter \fi\fi
  \bgroup \let\par\relax
  \let\@sharp##\let\protect\relax
  \@arrayskip\@preamble}
\def\eqnarray{\stepcounter{equation}%
              \let\@currentlabel=\theequation
              \global\@eqnswtrue
              \global\@eqcnt\z@
              \tabskip\@centering
              \let\\=\@eqncr
 \halign to \displaywidth\bgroup
    \eqnumphantom\@eqnsel\hskip\@centering
    $\displaystyle \tabskip\z@ {##}$%
    \global\@eqcnt\@ne \hskip 2\arraycolsep
         $\displaystyle\arraymode{##}$\hfil
    \global\@eqcnt\tw@ \hskip 2\arraycolsep
         $\displaystyle\tabskip\z@{##}$\hfil
         \tabskip\@centering
    &{##}\tabskip\z@\cr}
\begingroup\ifx\undefined\newsymbol \else\def\input#1 {\endgroup}\fi
\newfont{\hr}{msbm10}
\newfont{\ams}{msam10}

\font\numbers=cmss12
\font\upright=cmu10 scaled\magstep1
\def\stroke{\vrule height8pt width0.4pt depth-0.1pt}
\def\topfleck{\vrule height8pt width0.5pt depth-5.9pt}
\def\botfleck{\vrule height2pt width0.5pt depth0.1pt}
\def\Zmath{\vcenter{\hbox{\numbers\rlap{\rlap{Z}\kern 0.8pt\topfleck}\kern
2.2pt
                   \rlap Z\kern 6pt\botfleck\kern 1pt}}}
\def\Qmath{\vcenter{\hbox{\upright\rlap{\rlap{Q}\kern
                   3.8pt\stroke}\phantom{Q}}}}
\def\Nmath{\vcenter{\hbox{\upright\rlap{I}\kern 1.7pt N}}}
\def\Cmath{\vcenter{\hbox{\upright\rlap{\rlap{C}\kern
                   3.8pt\stroke}\phantom{C}}}}
\def\Rmath{\vcenter{\hbox{\upright\rlap{I}\kern 1.7pt R}}}
\def\Z{\ifmmode\Zmath\else$\Zmath$\fi}
\def\Q{\ifmmode\Qmath\else$\Qmath$\fi}
\def\N{\ifmmode\Nmath\else$\Nmath$\fi}
\def\C{\ifmmode\Cmath\else$\Cmath$\fi}
\def\R{\ifmmode\Rmath\else$\Rmath$\fi}

\newcounter{app}

\def\app{\setcounter{equation}{0}
\def\theequation{\Alph{app}.\arabic{equation}}\par
   \addvspace{4ex}
   \@afterindentfalse
  \secdef\@app\@dapp}

\newcommand\@app{\@startsection {app}{1}{0ex}%
                                   {-3.5ex \@plus -1ex \@minus -.2ex}%
                                   {2.3ex \@plus.2ex}%
                                   {\normalfont\Large\bf}}
\def\@dapp#1{%
{\parindent \z@ \raggedright  \bf #1}\par\nobreak}
\def\l@app#1#2{\ifnum \c@tocdepth >\z@
    \addpenalty\@secpenalty
    \addvspace{1.0em \@plus\p@}%
    \setlength\@tempdima{8em}%
    \begingroup
      \parindent \z@ \rightskip \@pnumwidth
      \parfillskip -\@pnumwidth
      \leavevmode \bfseries
      \advance\leftskip\@tempdima
      \hskip -\leftskip
      #1\nobreak\hfil \nobreak\hb@xt@\@pnumwidth{\hss #2}\par
    \endgroup\fi}
\newcounter{sapp}[app]

\def\sapp{\def\theequation{\Alph{app}.\arabic{equation}}
\par
\@afterindentfalse
  \secdef\@sapp\@dsapp}
\newcommand{\@sapp}{\@startsection{sapp}{2}{\z@}%
                                     {-3.25ex\@plus -1ex \@minus -.2ex}%
                                     {1.5ex \@plus .2ex}%
                                     {\normalfont\large\bfseries}}

\def\@dsapp#1{%
{\parindent \z@ \raggedright  \bf #1
}\par\nobreak}
\newcommand{\l@sapp}{\@dottedtocline{2}{1.5em}{2.3em}}

\def\Bf#1{\mbox{\boldmath $#1$}}
\def\balpha{{\Bf\alpha}}

\def\bgamma{{\Bf\gamma}}

\def\d{\partial}

\def\2{{1\over 2}}
\def\N2{${\cal N}=2$}
\def\4N{${\cal N}=4$}
\def\1N{${\cal N}=1$}

\def\be{ \begin{eqnarray} }
\def\ee{ \end{eqnarray} }
\def\d{\partial}
\def\bea{\begin{eqnarray}}
\def\eea{\end{eqnarray}}

\def\beq{\begin{equation}}
\def\eeq{\end{equation}}
\def\ba{\beq\new\begin{array}{c}}
\def\ea{\end{array}\eeq}
\def\be{\ba}
\def\ee{\ea}

\begin{document}


\begin{flushright}
FIAN/TD-26/99\\
ITEP/TH-78/99\\
hep-th/9912124
\end{flushright}
\vspace{0.5cm}
\begin{center}
{\LARGE \bf Duality in Integrable Systems and Generating Functions
for New Hamiltonians}\\
\vspace{0.5cm}
{\Large A.Marshakov
\footnote{e-mail address: mars@lpi.ru,\ andrei@heron.itep.ru}}\\
\vspace{0.5cm}
{\it Theory Department, Lebedev Physics Institute, Moscow
~117924, Russia\\ and \\ ITEP, Moscow ~117259, Russia},
\\
\end{center}
\bigskip
\begin{quotation}
Duality in the integrable systems arising in the context of
Seiberg-Witten theory shows that their tau-functions indeed can be seen
as generating functions for the mutually Poisson-commuting hamiltonians of
the {\em dual} systems. We demonstrate that the $\Theta$-function
coefficients of their expansion can be expressed entirely in terms of
the co-ordinates of the Seiberg-Witten
integrable system, being, thus, some set of hamiltonians for a dual system.
\end{quotation}

\section{Introduction}

It has become clear recently that rather simple and well-known
finite-dimensional {\em integrable} hamiltonian systems play an
important role in the formulation of effective action for low-dimensional
SUSY string and Yang-Mills theories \cite{GKMMM} (see also \cite{M} and
list of references therein). Remarkably enough, the
nonlinear dynamics arises on moduli spaces of vacuum expectation values
of scalar fields and Wilson loops of gauge fields, which play the role
of dynamical variables (co-ordinates, momenta and action-angle variables) of
an integrable system, which usually belongs to a relatively wide class
of (complexified) "systems of particles" of Toda, Calogero-Moser and
Ruijsenaars-Schneider type (the most recent "classification" of the
Seiberg-Witten \cite{SW,DW} integrable systems can be found in \cite{BMMM2}).

Speaking about this family of integrable models,
Seiberg-Witten theory, on one hand, explains the physical origin of
"relativization" of integrable systems of particles in spirit of S.Ruijsenaars
-- the contribution of the Kaluza-Klein sector in theories with extra
compact dimensions \cite{Nek}. On the other hand, it considers co-ordinates
and momenta on equal footing with the action-angle variables \cite{M99} -- as
moduli in compactified theory \cite{SW3}. The general picture of the
Seiberg-Witten theory from this perspective and general ideology of duality
in modern string or M-theory leads to expecting of some duality
relations between co-ordinates and momenta on one side and action-angle
variables on another one. It is not a coincidence thus, that such duality was
found to act naturally on the phase spaces of Calogero-Moser models and their
"compact" Ruijsenaars-Schneider relatives \cite{dual,BMMM3} (the Toda chain
systems arise as a special "double-scaling" degenerate case \cite{Ino}).

This duality can be considered as a way of constructing new nontrivial
completely integrable systems with very nontrivial properties. For example,
it may lead to the systems where the hamiltonians depend on momenta (or on
both momenta and co-ordinates) through the elliptic double-periodic functions
(next step after the Ruijsenaars "relativization" when hamiltonians depend
on momenta via trigonometric functions), an example of such double-elliptic
system was proposed in \cite{BMMM3}. It was also conjectured in \cite{BMMM3}
that the tau-functions of the Seiberg-Witten finite gap integrable systems
(expressed through theta functions on corresponding Seiberg-Witten curves
with specific period matrices -- the set of couplings in low-energy
effective SUSY gauge theories) may play a role of generating functions for the
hamiltonians of the dual systems. Recently this conjecture was checked for
the "perturbative" or "instanton" expansions by numeric MAPLE computations
\cite{mim}. Below in this note we will demonstrate that mutual Poisson
commutativity of hamiltonians follows directly from duality arguments --
the $\Theta$-coefficients of the expansion of generating function are
functions only of co-ordinates (or dual action variables) of the original
Seiberg-Witten integrable system, or the function of moduli entering the
superpotential of compactified theory \cite{SW3,compsup,Dorey,M99}.

\section{Duality in Integrable Systems}

By duality for a system of {\em free} particle one usually means the Fourier
transform, or exchange between co-ordinates and momenta, which are the
integrals of motion themselves in free case. One may think of an
exchange between the co-ordinates and {\em action} variables as of
a natural generalization of the Fourier transform for a nontrivial or
interacting dynamical system, if it is completely integrable, since both
set of quantities look similar from the hamiltonian point of view --
i.e. mutually Poisson commute.  Following \cite{dual, BMMM3}, by {\em dual}
systems we call integrable systems "living" on the same phase space, with the
same symplectic form
\be
\label{symp}
\Omega = d{\bf q}\wedge d{\bf p} = d{\bf a}\wedge d{\bf z}
\ee
but with two {\em different} set
of integrals of motion (action variables or hamiltonians which are the
functions of each other) -- for the first (the "Seiberg-Witten") system the
hamiltonians are $\{ h_i({\bf q}, {\bf p}) \}$ (or action variables $\{
a_i({\bf q}, {\bf p}) \}$, $h_i = h_i({\bf a})$) -- some nontrivial functions
of co-ordinates and momenta, while for the dual system the role of action
variables (or hamiltonians) is played by the original co-ordinates (or some
algebraically independent functions of them) $\{ q_i({\bf a}, {\bf z}) \}$.
Both sets mutually Poisson commute \be \{ q_i, q_j \} = 0 \ \ \ \ \ \ \ \ \ \
\{ a_i, a_j \} = 0
\ee
$i,j=1,\dots,\2\dim({\rm Phase\ Space})$, with respect to the Poisson bracket
inverse to (\ref{symp}).

For arbitrary integrable systems this very general property (defined already
at the level of the Liouville theorem) does not lead necessarily to any
interesting consequences. However, the Seiberg-Witten integrable systems
possess very peculiar properties with respect to the duality transformation
${\bf q}\leftrightarrow{\bf a}$. In the perturbative (weak coupling) limit,
Seiberg-Witten models with the adjoint matter are described by trigonometric
Calogero and Ruijsenaars models, the family of which is {\em self-dual} with
respect to duality transformation \cite{dual}, i.e. the {\em functions} $a_k
= f_k({\bf q},{\bf p})$ and $q_k = f_k({\bf a},{\bf z})$, expressing the action
variables via co-ordinates and momenta and the functions expressing "new"
actions ("old" co-ordinates) via "new" co-ordinates and momenta ("old"
action and angle variables) {\em coincide}, i.e. the duality transformation
is "symmetric". This is literally true for the trigonometric Ruijsenaars and
rational Calogero models (with the hamiltonians $h = 2\cosh p\sqrt{1 + {m^2\over
\sinh^2q}}$ and $h = p^2 + {m^2\over q^2}$ -- in the simplest case --
respectively; one may easily chack this solving directly equations of motion
in one-particle case) while the trigonometric Calogero model is dual to the
rational Ruijsenaars.

This family of integrable models is nothing but a perturbative degeneration
of the finite-gap or Hitchin integrable systems corresponding to the
Seiberg-Witten curves $\Sigma$ of the Lax form
\be
\label{laxcu}
\det (\lambda -  {\cal L}(z)) = 0
\ee
where the Lax operator ${\cal L}(z)$ is defined on some {\em base} curve
$\Sigma_0$ -- below (and usually) torus (elliptic) or sphere with punctures
(rational) -- with generating differential
\be
\label{dS}
dS = \lambda dz
\\
\delta_{\rm moduli} dS = {\rm holomorphic}
\ee
The action variables are given by the Seiberg-Witten contour
integrals over half of the independent contours
\be
{\bf a} = \oint_{\bf A}dS
\ee
or
\be
{\bf a}^D = \oint_{\bf B}dS
\ee
and the symplectic form is \cite{DKN,KriPho,M96}
\be
\Omega = \left.\delta dS\right|_{\bgamma} = d{\bf a}\wedge d{\bf z}(\bgamma)
= d{\bf q}\wedge d{\bf p}
\ee
where the variation of differential (\ref{dS}) is computed in the divisor
${\bgamma}$ of the poles of the Baker-Akhiezer function.

In the perturbative limit the curve (\ref{laxcu}) becomes rational, what
corresponds to the solitonic limit of a finite-gap system. Then dependence
on base curve (on spectral parameter $z$) effectively disappears and
eq.~(\ref{laxcu}) becomes a generating function for the hamiltonians of
(perturbative) Seiberg-Witten integrable system -- invariant combinations
of matrix elements of ${\cal L}$
\be
\label{laxdege}
\det(\lambda - {\cal L}) = \sum \lambda^kh_k
\ee
On the other hand, the determinant in (\ref{laxdege}) looks very similar
to the solitonic or wronskian {\em tau-functions} for the solutions to KP or
Toda lattice hierarchies, if the spectral parameter $\lambda$ would play
the role of "main" (first in KP case or zero in Toda) time -- always a
parameter on base curve. This way the particular analytic
representation of the Seiberg-Witten curve could be thought of
as a tau-function for a {\em dual} integrable system, or, better, of a
solution to KP or Toda lattice hierarchy associated with a dual system in the
sense of \cite{AMcM,KriCal,KriZa} -- when the zeroes of tau function of
KP/Toda hierarchy become co-ordinates of some integrable system of particles.
The discussion of the parallels between tau-functions and spectral curve
equations (\ref{laxcu}) is beyond the scope of this note, it only helps now
to state that it could be natural to look for the generation function of dual
mutually Poisson-commuting hamiltonians in terms of tau-functions of the
Seiberg-Witten finite gap solutions -- the theta functions on Seiberg-Witten
spectral curves.

\section{Generating Function}

Let us turn now directly to the generating functions \cite{BMMM3} to be
discussed below.  We will consider the situation when the spectral curve
(\ref{laxcu}) covers 1-dimensional complex torus, or its degeneration -- a
cylinder or sphere with two punctures. The last case corresponds, among
others, to the periodic Toda chain or pure glyodynamics while the first case
contains the elliptic Calogero-Moser and Ruijsenaars-Schneider models,
corresponding to the Seiberg-Witten theories with adjoint matter.

Consider the decomposition of the Riemann
theta-function, defined on Jacobian of a curve of genus $g=N$
\be
\label{theta}
\Theta ({\bf z}|T) = \sum_{{\bf n}\in{\Z }^N}
e^{2\pi i\sum_{i=1}^Nn_iz_i + i\pi\sum_{i,j=1}^Nn_iT_{ij}n_j}
\ee
when period matrix satisfies the constraint
\be
\label{matrco}
\sum_{i=1}^N T_{ij}\stackrel{\forall j}{=}\ \tau\ \stackrel{\forall
i}{=}\sum_{j=1}^NT_{ij}
\ee
naturally arising in "elliptic" models with base torus. In other terms
\be
T_{ij} = {\tau\over N} + {\tilde T}_{ij}
\\
\sum_{i=1}^N {\tilde T}_{ij}\stackrel{\forall j}{=}\ 0\ \stackrel{\forall
i}{=}\sum_{j=1}^N{\tilde T}_{ij}
\ee
Introducing also $ z_i = {z\over N} + {\tilde z}_i$ with $\sum_{i=1}^Nz_i=z$
or $\sum_{i=1}^N{\tilde z}_i=0$, one immediately gets
\be
\label{dec}
\Theta ({\bf z}|T) = \sum_{{\bf n}\in{\Z }^N}
e^{2\pi i{z\over N}\sum_{i=1}^Nn_i + i\pi{\tau\over
N}\left(\sum_{i=1}^Nn_i\right)^2 +
2\pi i\sum_{i=1}^Nn_i{\tilde z}_i + i\pi\sum_{i,j=1}^Nn_i{\tilde T}_{ij}n_j}
=
\\
= \sum_{ k\in {\Z }}e^{2\pi i{k\over N}z + i\pi {k^2\over N}\tau}
\sum_{{\bf n}\in{\Z }^N; \sum_{i=1}^N n_i=k}
e^{2\pi i\sum_{i=1}^Nn_i{\tilde z}_i + i\pi\sum_{i,j=1}^Nn_i{\tilde T}_{ij}n_j}
\ee
Presenting $k=Nm + i$ with $m\in{\Z }$ and $i\in {\Z }_N = {\Z }\ {\rm
mod} N$ (i.e. $i=0,1,\dots,N-1$), one finally gets \cite{BMMM3}
\be\label{decfin}
\Theta ({\bf z}|T) = \sum_{i\in{\Z }_N}\sum_{m\in{\Z }}
e^{2\pi i\left(m+{i\over N}\right)z + i\pi N\tau\left(m+{i\over N}\right)^2}
\sum_{{\bf n}\in{\Z }^N; \sum_{j=1}^N n_j=i}
e^{2\pi i\sum_{j=1}^Nn_j{\tilde z}_j + i\pi\sum_{j,j'=1}^Nn_j{\tilde T}_{ij}
n_{j'}} =
\\
= \sum_{i\in{\Z }_N}\theta_{i\over N}(z| N\tau)\Theta_{i}({\bf\tilde z}|
{\tilde T})
\ee
where $\theta_{i\over N}(z| N\tau)$ is genus $g=1$ or Jacobi theta-functions
with specific characteristics ${i\over N}\equiv\left[^{i\over N}_{\ 0}\right]$,
while $\Theta_{i}({\bf\tilde z}|{\tilde T})$ is defined on Jacobian of genus
$g=N-1$. Indeed, for example it can be rewritten as
\be
\label{thetahat}
\Theta_{i}({\bf\tilde z}|{\tilde T}) =
\sum_{{\bf n}\in{\Z }^N; \sum_{j=1}^N n_j=i}
e^{2\pi i\sum_{j=1}^Nn_j{\tilde z}_j + i\pi\sum_{j,j'=1}^Nn_j{\tilde T}_{ij}
n_{j'}} =
\\
= \sum_{{\bf m}\in\left({\Z }-{i\over N}\right)^{N-1}}
e^{2\pi i\sum_{j=1}^{N-1}m_j{\hat z}_j + i\pi\sum_{j,j'=1}^{N-1}m_j
{\hat T}_{ij}m_{j'}} \equiv \Theta_{i}({\bf\hat z}|{\hat T})
\ee
with
\be
{\hat z}_j = {\tilde z}_j - {\tilde z}_N = {\tilde z}_j + \sum_{l=1}^{N-1}
{\tilde z}_l
\\
{\hat T}_{ij} =
{\tilde T}_{ij} + \sum_{k=1}^{N-1}({\tilde T}_{ik} + {\tilde T}_{kj}) +
\sum_{k,l=1}^{N-1}{\tilde T}_{kl}
\ee
Our aim is to study the properties of the decomposition
(\ref{decfin}) and to demonstrate that the ratios of the coefficients
$\Theta_i$ are functions only of the co-ordinates of the integrable systems
thus being a set of independent hamiltonians for a dual system.

\subsection{Toda chain}

Consider, first, a "degenerate" case of the periodic Toda chain with the
base curve $\Sigma_0$ being a cylinder. If one substitutes into
(\ref{thetahat}) the period matrix of the
genus $g=N-1$ periodic Toda chain spectral curve (\ref{laxcu})
\be
\label{todacu}
w + {\Lambda^{2N}\over w} = P_N(\lambda) = \lambda^N + \sum_{k=0}^{N-2}
h_k\lambda^k
\ee
with the Seiberg-Witten integrals (on cylinder it is
natural to choose $z=\log w$) \be a_i = \oint_{A_i}\lambda{dw\over w} \\
a^D_i = \oint_{B_i}\lambda{dw\over w}
\\
{\tilde T}_{ij} = {\d a^D_i\over\d a_j}
\ \ \ \ \ \ \ i,j=1,\dots,N-1
\ee
one gets that
\be\label{dectoda}
\Theta ({\bf z}|T) = \sum_{k\in{\Z }_N}
e^{2\pi i{k\over N}z} \Theta_{k}
\ee
and
\be
\label{poicom}
\left\{ {\Theta_i\over\Theta_j}, {\Theta_{i'}\over\Theta_{j'}}\right\} = 0
\\
\forall\ i,j,i',j'
\ee
where the Poisson bracket is taken w.r.t. symplectic form
\be
\label{sympfor}
\Omega^{Toda} = \sum_{i=1}^{N-1}d{\hat z}_i\wedge da_i =
\sum_{i=1}^{N-1}d{q}_i\wedge dp_i
\ee
since $\Theta_i\equiv\Theta_{i}({\bf\tilde z}|{\tilde T})$ describe exactly
the co-ordinartes of the solution to periodic Toda chain \cite{DaTa,KriTo}
\be
\label{todasol}
e^{q_i} = {\Theta_i\over\Theta_{i-1}}
\\
{\Theta_i\over\Theta_j} = \prod_{k=j+1}^i e^{q_k}
\ee
and obviously commute ($\{ q_i,q_j\} = 0$, $\forall i,j$) w.r.t.
(\ref{sympfor}) since \cite{KriPho,M96}
\be
\label{sympfor1}
\Omega^{Toda} = \sum_{i=1}^{N-1}d{\hat z}_i\wedge da_i =
\sum_{i=1}^{N-1}dq_i\wedge dp_i
\ee
where $q_i$ and $p_i$ are co-ordinates and momenta of the Toda chain
particles. $\Theta_i$ (\ref{todasol}) could be thought of as the Toda chain
tau-functions, depending on discrete time $i$ -- the number of particle.
Eq.~(\ref{poicom}) gives an exact form of old expectation that the Toda
chain tau-functions Poisson commute with each other. In the perturbative
limit $\Lambda\to 0$ spectral curve (\ref{todacu}) degenerates into rational
$w = P_N(\lambda )$, the period matrix becomes $i\pi{\tilde T}_{ij} =
{\d^2\over\d a_i\d a_j}\2\sum_{k<l}^N(a_k-a_l)^2\log{a_k-a_l\over\Lambda}$
and $\Theta_k$, after redefining ${\tilde z}_i\to{\tilde z}_i-{N\over 2\pi i}
\log\Lambda$, $\Theta_k\to\Lambda^{k^2}\Theta_k$ turn into expressions for the
tau-functions of open Toda chain.

\subsection{Calogero-Moser model}

In the Calogero-Moser case the
commutativity of $\Theta_i$ can be shown in the following way.
The (genus $g=N$) spectral curve is given by
\be\label{LaxCal}
\det_{N\times N}\left({\cal L}^{CM}(z) - \lambda\right) = 0
\\
{\cal L}^{CM}(z) =
\left({\bf pH} + \sum_{\balpha}F({\bf q\balpha}|z)
E_{\balpha}\right)
\\
F(q|z) = m\frac{\theta_{\ast}(q+z|\tau )}{\theta_{\ast}(q|\tau )
\theta_{\ast}(z|\tau )}e^{\zeta(q|\tau )z}
\ee
where $\theta_{\ast}(z)$ is odd Jacobi theta-function, and
\be
a_i = \oint_{A_i}\lambda dz
\\
a^D_i = \oint_{B_i}\lambda dz
\\
T_{ij} = {\d a^D_i\over\d a_j}
\\
i,j=1,\dots,N
\ee
so that eq.~(\ref{decfin}) can be considered literally. According to
\cite{KriCal} equation
\be
0 = \Theta ({\bf z}|T)
= \sum_{i\in{\Z }_N}\theta_{i\over N}(z| N\tau)\Theta_{i}
\ee
as an equation on $z$-torus has exactly $N$ zeroes ${1\over N}z=q_1,\dots,q_k$. As a
result one gets a system of linear equations
\be
\label{systhet}
\sum_{i=1}^N\theta_{i\over N}(Nq_j| N\tau)\Theta_{i} = 0
\ \ \ \ \ \ j=1,\dots,N
\ee
The system should have nontrivial solutions, i.e.
$\det_{ij}\theta_{i\over N}(Nq_j| N\tau) = 0$, which effectively reduces the
number of degrees of freedom from $N$ to $N-1$.
Then (\ref{systhet}) can be
rewritten as
\be
\sum_{i\neq i_0}^{N-1}\theta_{i\over N}(Nq_j| N\tau)
{\Theta_{i}\over\Theta_{i_0}} = \theta_{i_0\over N}(Nq_j| N\tau)
\\
\forall i_0; \ \ \ \ \ j=1,\dots,N-1
\ee
i.e.
\be
\label{sollisy}
{\Theta_{i}\over\Theta_{i_0}} = {\det_{k\neq i_0,i\to i_0;j=1,\dots,N-1}
\theta_{k\over N}(Nq_j| N\tau)\over\det_{k\neq i_0;j=1,\dots,N-1}
\theta_{k\over N}(Nq_j| N\tau)}
\ee
Therefore, the ratios ${\Theta_{i}\over\Theta_{j}}$ depend only on the
co-ordinate $q_k$, $k=1,\dots,N$ of the Calogero-Moser particles and, thus,
obviously Poisson commute $\left\{ {\Theta_i\over\Theta_j},
{\Theta_{i'}\over\Theta_{j'}}\right\} = 0$, $\forall\ i,j,i',j'$ with
respect to the Calogero-Moser symplectic structure
\be
\label{sympfor2}
\Omega^{CM} = \sum_{i=1}^{N}d{q}_i\wedge dp_i
= \sum_{i=1}^{N}d{z}_i\wedge da_i
\ee
restricted to $\sum_{j=1}^Nq_j = const$ (the condition of vanishing
the determinant $\det_{ij}\theta_{i\over N}(Nq_j| N\tau) = 0$) and
\be
\sum_{j=1}^Na_j = \oint_{\sum_{j=1}^NA_j}\lambda dz =
const\cdot\oint_{A_0}dz = const
\ee
(where $A_0$ is A-cycle on base curve $\Sigma_0$).
The explicit form of the solutions (\ref{sollisy}) can be easily found using
the $\theta$-function identities, coming from the Wick theorem \cite{Wick}
\be
\label{wick}
\det_{ij}\theta_{i\over N}(Nq_j| N\tau) \sim
\theta_{\Sigma {i\over N}}\left(N\sum_k q_k | N\tau\right)
\prod_{i<j}\theta_{\ast}(Nq_i-Nq_j| N\tau)
\ee
so that the (squared) solutions (\ref{sollisy}) become elliptic functions on
base torus $\Sigma_0$.

\subsection{Generalities}

Similar arguments can be applied in the "relativistic" case of the elliptic
Ruijsenaars-Schneider model. The spectral curve $\Sigma$ is again defined
over elliptic base $\Sigma_0$ (see, for example, \cite{BMMM2} and references
therein for details) and one can study zeroes of, now, Toda lattice
tau-function in double-periodic variable -- the zero (or discrete) time
\cite{KriZa}.  The formulas of the previous section are generalized
straightforwardly.

The situation is not so simple with hypothetic double-elliptic system for
the general case of $N$ degrees of freedom. The only known example exists
for the $N=1$ case \cite{BMMM3} and it is self-dual with respect to ${\bf
q}\leftrightarrow{\bf a}$ exchange. There is no clear way yet to write down
any reasonable Lax representation or spectral curve equation, except for the
perturbative case, when it can be identified with $w = w_0\cdot
{\prod\theta_{\ast}(\xi - \xi_i)\over\theta_{\ast}(\xi - \eta_i)}$ with
$dS = \xi d\log w$ and is a degenerate case of the $XYZ$ spin chain
\cite{mami97}. This problem deserves further investigation.

\section{Discussion}

In this note we have discussed the conjecture of \cite{BMMM3} that
tau-functions of the Seiberg-Witten integrable systems can play a role of
generating functions for new commuting hamiltonians. As a result it is
demonstrated above, that the arising quantities -- the ratios of coefficients
of decomposition over base curve are indeed Poisson-commuting quantities --
the (functions of) hamiltonians of dual integrable systems or, in other
words, the can be expressed via only the co-ordinates of the original
integrable system.

Being an interesting academic question from the internal point of view of
the theory of integrable systems, it is also related to more physical issues
of the Seiberg-Witten theory. Co-ordinates and momenta of integrable system
play the role of bare moduli in Seiberg-Witten theory so that
the co-ordinates are the variables associated with superpotential in
compactified theory with partially broken supersymmetry. It would be
interesting to understand the role of generating function and $\{ \Theta_i\}$
from general point of view of geometric description of the
Seiberg-Witten M-theory vacua.  We hope to return to this problems elsewhere.

\section*{Acknowledgements}

I am grateful to H.Braden, V.Fock, S.Kharchev, I.Krichever, A.Mironov and
A.Morozov for useful discussions. The research was partly supported by the
RFBR grant 98-01-00344 and INTAS grant 97-0103.

\end{document}